 \documentstyle{laa}     
\def\ut#1{\mathop{\vtop{\ialign{##\crcr
     $\hfil\displaystyle{#1}\hfil$\crcr\noalign
     {\kern1pt\nointerlineskip}\hbox{$\hfil\sim\hfil$}\crcr
     \noalign{\kern1pt}}}}}

\def\undersymbol#1#2{\mathop{\vtop{\ialign{##\crcr
     $\hfil\displaystyle{#2}\hfil$\crcr\noalign
     {\kern1pt\nointerlineskip}\hbox{$\hfil#1\hfil$}\crcr
     \noalign{\kern1pt}}}}}
\def\arcsec{^{\prime\prime}}
\def\arcmin{^{\prime}}

\begin{document}
\thesaurus{03.12.2 - 07.30.1 - 19.31.3 - 24.03.1}
\title{X-ray emission from dark clusters of MACHOs}
\author{
    F. De Paolis   \inst{1} 
\and G. Ingrosso   \inst{2} 
\and  Ph. Jetzer   \inst{1}
\and M. Roncadelli \inst{3} }
\offprints{Ph. Jetzer}
\institute{
Paul Scherrer Institute, Laboratory for Astrophysics, CH-5232 Villigen PSI, and
Institute of Theoretical Physics, University of Zurich, Winterthurerstrasse
190, CH-8057 Zurich, Switzerland
\and
Dipartimento di Fisica, Universit\`a di Lecce, and INFN, Sezione di Lecce, 
Via Arnesano, CP 193, I-73100 Lecce, Italy
\and
INFN, Sezione di Pavia, Via Bassi 6, I-27100, Pavia, Italy}
\date{Received date; accepted date}
\maketitle
\begin{abstract}
MACHOs (Massive Astrophysical Compact Halo Objects) -
as discovered by microlensing experiments towards the LMC -
provide a natural explanation for the galactic halo dark matter.
A realistic possibility is that MACHOs are brown dwarfs of mass 
$\sim 0.1 M_{\odot}$. Various arguments suggest that brown dwarfs
should have a coronal X-ray emission of $\sim 10^{27}$ erg s$^{-1}$. 
As MACHOs are presumably clumped into dark clusters (DCs),
each DC is expected to have a total X-ray luminosity 
of $\sim 10^{29}-10^{32}$ erg s$^{-1}$.
We discuss the possibility that dark clusters contribute to the diffuse
X-ray background (XRB) or show up as discrete sources in very deep
field X-ray satellite observations. 
Moreover, from the observed diffuse XRB
we infer that the amount of virialized diffuse gas present in the galactic 
halo can at most make up $5 \%$ of the halo dark matter.
\end{abstract}
 
\section{Introduction}
In spite of the dynamical evidence that a lot of dark matter is 
concealed in galactic halos,
observations failed for many years to provide any information about its 
nature.
A breakthrough came recently with the discovery of 
Massive Astrophysical Compact Halo Objects (MACHOs) in microlensing 
experiments towards the Large Magellanic Cloud (Alcock et al. \cite{alcock}; 
Aubourg et al. \cite{aubourg}). 
Today, although the existence of MACHOs -- presumably located in 
the halo of our galaxy -- appears to be firmly established,
the implications of such an achievement are still largely unsettled, owing
to their sensitive dependence on the assumed galactic model 
(Gates, Gyuk \& Turner \cite{ggt}).
More specifically, the existing data-set permits to 
reliably conclude that MACHOs should lie
in the mass range $0.05~M_{\odot} - 1.0~M_{\odot}$,
but stronger claims are unwarranted because the inferred MACHO mass strongly
depends on the uncertain properties of the considered galactic model 
(Evans \cite{evans}, 
De Paolis, Ingrosso \& Jetzer \cite{dij}).

Even if present uncertainties do not yet permit to make any sharp statement
about the nature of MACHOs, brown dwarfs look as a viable
possibility to date, and we shall stick to it throughout.

Remarkably enough, a simple scenario has been proposed 
(De Paolis et al. \cite{depaolis2}; 
Gerhard \& Silk \cite{gerhard};  
Nulsen and Fabian \cite{nulfab}) which -- besides encompassing the
Fall-Rees theory for the formation of globular clusters (Fall \& Rees
\cite{fall}) -- predicts that dark clusters of brown dwarfs and cold
molecular clouds (mainly of $H_2$) should form in the halo at
galactocentric distances larger than 10--20 kpc. Dark  clusters in the
mass range $3\times 10^2 - 10^6~M_{\odot}$ are expected to have
survived all disruptive effects and should still populate the outer
galactic halo today (De Paolis et al. \cite{vangelo,apocalisse}).

In this paper we would like to discuss  further aspects of the above
scenario, which naturally arise provided MACHOs 
have coronal X-ray emission. Indeed, 
X-ray observations of field stars 
have shown 
that low-mass stars -- with mass even as low as $0.08~M_{\odot}$ --
do have coronal X-ray emission. Thus -- given that MACHOs are old,
very low-mass stars -- we expect them to be X-ray active. Actually, under 
this assumption, Kashyap et al. \cite{kashyap} have explored 
the possibility that an {\it unclustered} MACHO 
population in the galactic halo 
can contribute significantly (at levels $> 10\%$) to the diffuse 
X-Ray Background (XRB) at higher ($\ge 0.5$ keV) energies. 
As we shall see in Sect. 4, such an unclustered distribution looks however
unplausible. Apart from automatically avoiding this difficulty, the 
scenario under consideration has two further implications as far as X-ray 
emission is concerned.
First, the gas present in the DCs partially
shields the X-ray emission from MACHOs, thereby reducing the total 
contribution to the XRB (as compared with the expectation of Kashyap et 
al. \cite{kashyap}). Second -- besides contributing to the diffuse XRB --
DCs can also show up as discrete sources in analyses of very deep field 
X-ray satellite 
observations towards regions in the sky of very low column density.

In fact, the existence of a 
new population of not yet fully recognized X-ray sources has been 
suggested by an analysis of very deep ROSAT observations
(towards the {\it Lockman Hole}), which 
revealed a considerable excess of faint X-ray sources over that expected 
by Active Galactic Nuclei (AGN) evolution models (Hasinger et al. 
\cite{hasinger}). A re-analysis of the same data 
shows that most of the sources in excess may be identified 
with clusters of galaxies and narrow emission or absorption line
galaxies (Hasinger \cite{hasinger_compton}).
Recently, data analyses of ROSAT observations in a different region 
of the sky (McHardy et al. \cite{mch}) have
confirmed the previous conclusions by Hasinger \cite{hasinger_compton} 
and make even more clear that below a flux level of 
$\sim 3 \times 10^{-15}$ erg cm$^{-2}$ s$^{-1}$ there exists a population of  
unidentified, faint point-like X-ray sources (about 10-20\% of the total).
Still, an open problem is whether the unidentified sources are
galactic or extragalactic in origin.  
Our main point is that a galactic population of DCs 
can significantly contribute 
to the inferred new population in question.

The plan of the paper is as follows.
In Section 2  we discuss the questions concerning the DCs 
X-ray emission, while the observational implications of the 
scenario under consideration are addressed in Section 3.
Limits on the amount of the virialized X-ray emitting diffuse gas 
in the galactic halo are derived in Section 4. Our conclusions are 
summarized in Section 5.

\section{Dark Clusters and X-ray Emission}
As is well known, low-mass stars 
have coronal X-ray emission 
due to the presence of collisionally excited plasma 
in their convective envelopes.
Observations have shown that the coronal gas temperature $T$ lies 
in the range  $\sim 10^6-10^8$ K (Schmitt et al. \cite{schmitt}).
As regard to the corresponding 
X-ray luminosity, measurements performed by EINSTEIN and ROSAT satellites
yield the average value $L_X\sim 10^{27}$ erg s$^{-1}$ for late
M-stars
(Barbera et al. \cite{barbera}, Fleming et al. \cite{fleming}).
Moreover, neither any obvious correlation between
$L_X$ and bolometric luminosities for K and M-stars has been found
(Mullan and Fleming \cite{mullan}), nor a dependence of $L_X$ on the 
metallicity has been detected (Kashyap et al. \cite{kashyap}).
Now, brown dwarfs are very low-mass stars which share several features 
in many spectral bands with late M-stars.
So, it seems plausible to expect MACHOs to have coronal X-ray emission as 
well 
\footnote{
The X-ray emission from lower mass objects (such as Jupiter-sized planets) is 
ignorable for our purposes, owing to their low 
X-ray luminosity $\sim 10^{16}-10^{17}$ erg s$^{-1}$.}. 
More specifically, we assume that MACHOs have an average X-ray luminosity 
$L_X^{~~M} \sim 10^{27}$ erg s$^{-1}$ in the $0.1 - 10$ keV energy band.
Our attitude is further supported by the direct observation of at least 
one object, the 
VB 8 dwarf (of mass $\simeq 0.08~M_{\odot}$), by EINSTEIN,  
EXOSAT and ROSAT satellites between 1979 and 1994, whose 
measured  X-ray luminosity
is  $1.8\times 10^{28}$ erg $s^{-1}$,
$2.6\times 10^{27}$ erg $s^{-1}$ and $8.3\times 10^{26}$ erg $s^{-1}$
in the energy bands $0.05-2$ keV, $0.12-3.7$ keV and $0.1-2.4$ keV, 
respectively (Drake et al. \cite{drake}). 

Let us focus our attention on the coronal plasma emissivity 
$\epsilon_{\nu}(\nu,T,Z) = f(\nu,T,Z) ~ n_i n_e$,
where $Z$ denotes the gas metallicity, 
$n_i$ and $n_e$ are the ion and electron number density, respectively (we
shall use cgs units throughout the paper).
Following Raymond and Smith \cite{rs}, 
this quantity is obtained by adding 
the continuum energy emissivity 
(by free-free and free-bound processes)
to the emission-line contribution by metals 
(through bound-bound processes). 
We may confidently suppose that the primordial halo gas from which MACHOs 
have formed has a metallicity $Z \sim 10^{-2}Z_{\odot}$
(from now on - with the exception of Section 4 - this value will be 
implicitly assumed throughout and the Z-dependence in the ensuing 
expressions will be dropped). 
Hence, the X-ray 
spectral luminosity $L_{\nu}^{~~M}(\nu,T)$ of a single MACHO is
obtained by integrating $\epsilon_{\nu}$ 
over the coronal volume $V_c$.
However, in the present 
instance, we do not need to know $V_c$, $n_i$ and $n_e$  
due to the obvious requirement
that integration of $L_{\nu}^{~~M}(\nu,T)$ over the frequency range
$0.1~{\rm keV}/h \leq \nu \leq 10~{\rm keV}/h$ is equal to the total
X-ray MACHO luminosity $L_X^{~~M}$.
In this way, we find
\begin{equation}
L_{\nu}^{~~M}(\nu,T) =
\frac {f(\nu,T)}
{\int_{0.1~{\rm keV}/h}^{10~{\rm keV}/h} d\nu f(\nu,T)} L_X^{~~M}~.
\end{equation}

Next, we compute the X-ray luminosity ${\cal L}_X^{~~DC}(\nu_1, \nu_2)$ 
of a DC in the frequency range $(\nu_1, \nu_2)$ taking 
the X-rays absorption by molecular clouds within the DC into account
(absorption by the galactic disk is momentarily disregarded).
To this end, it is convenient to consider first the X-ray luminosity
$I_X(b,\nu_1,\nu_2)$ in the frequency range
$(\nu_1,\nu_2)$ from a unit element of the DC projected surface --
perpendicular to the line of sight -- at impact parameter $b$.
We denote by $y$ the coordinate along the line of sight (with origin on 
the perpendicular plane through the centre of the DC), and by
$n_M$ and $n_{\rm mol}$ the number density of MACHOs and gas, respectively,
in a DC. In addition, $\sigma_X(\nu)$ stands for the X-ray absorption 
cross-section whereas $R_{DC}$ denotes the (median) radius of a DC.
It follows that the optical depth inside a DC is
\begin{equation}
\tau_{DC}(\nu, b,y) = 
\int_y^{\sqrt{R_{DC}^2-b^2}}dy'\sigma_X(\nu)~n_{\rm mol}~.
\end{equation}
Moreover, a straightforward calculation yields
\begin{equation}
\begin{array}{ll}
I_X(b,\nu_1,\nu_2) = \int_{\nu_1}^{\nu_2} d\nu ~ \times \\ \\ 
                     \int^{\sqrt{R^2_{DC} - b^2}}_{-\sqrt{R^2_{DC} - b^2}}
dy~ n_M~ L^{~M}_{\nu}(\nu,T)~ e^{-\tau_{DC}(\nu,b,y)}~.
\end{array}
\end{equation}
Because of the spherical symmetry, we ultimately get
\begin{equation}
{\cal L}_X^{~~DC}(\nu_1,\nu_2) = \int_0^{R_{DC}}
~I_X(b,\nu_1,\nu_2)~2\pi~ b~ db~.
\label{abs}
\end{equation}

What are $n_M$ and $n_{\rm mol}$?
Although the gas distribution is expected to be clumpy, for simplicity we 
can safely adopt a uniform average distribution, so that $n_{\rm mol}$ is 
constant. 
As discussed elsewhere (De Paolis et al. \cite{vangelo,apocalisse}), 
in the lack of any observational 
information about DCs it seems natural to take their 
average mass density equal to that of globular clusters. 
Then the following relationship ensues
$R_{DC} = 0.12~(M_{DC}/M_{\odot})^{1/3}$
($M_{DC}$ is the mass of a DC). 
On account of it, $n_{\rm mol}$ 
becomes independent of $M_{DC}$ and $R_{DC}$, and its explicit value is 
$\sim 10^3$ cm$^{-3}$.
The case of  $n_M$ is less trivial, since the results we are going to 
derive can depend
strongly on how MACHOs are distributed inside DCs. Below, we consider 
the two extreme situations:
($a$) all MACHOs are uniformly distributed in the DCs;
($b$) all MACHOs are concentrated in the DC cores. 
Realistically, we expect an intermediate situation to occur,
since the mass stratification instability drives the more massive
MACHOs into the cluster cores. 
In addition,  DCs with mass in the range 
$3\times 10^2~M_{\odot}\ut < M_{DC}\ut < 5\times 10^4~M_{\odot}$ should 
have started core collapse (De Paolis et al. \cite{apocalisse}), 
in which case situation ($b$) becomes an excellent approximation.

Let us proceed to address the two cases separately.
In case ($a$), $I_X(b,\nu_1,\nu_2)$ can be computed analytically to give
\begin{equation}
\begin{array}{ll}
I_X(b, \nu_1, \nu_2) = \displaystyle{\frac{n_M}{n_{\rm mol}}} 
\int_{\nu_1}^{\nu_2} d\nu~L^{~M}_{\nu}(\nu,T) ~ \times \\ \\
\displaystyle{
\frac{1-e^{-2\sigma_X(\nu)n_{\rm mol}\sqrt{R^2-b^2}}}{\sigma_X(\nu)}}~.
\end{array}
\label{4}
\end{equation}
By inserting expression (\ref{4}) into equation (\ref{abs}) we obtain 
\begin{equation}
{\cal L}_X^{~~DC}(\nu_1, \nu_2) =10 f \left(\frac{M_{DC}}{M_{\odot}}\right) 
L_X^{~~M}~g(\nu_1,\nu_2,T)~,
\label{7}
\end{equation}
where $f$ is the fraction of DC matter in the form of MACHOs,
$\beta\equiv\tau_{DC}(\nu, 0, 0)$ and we have introduced the attenuation
factor
\begin{equation}
g(\nu_1, \nu_2, T) =
\int_{\nu_1}^{\nu_2} \frac{3
L_{\nu}^{~M}(\nu, T)}{4 L_X^{~M}} 
\left[\frac{1}{\beta}+
\frac{e^{-2\beta}}{\beta^2}+
\frac{e^{-2\beta} - 1}{2\beta^3} \right] d\nu 
\label{8}
\end{equation}
which yields the fraction of X-rays emerging from a DC. 
Case ($b$) can be handled in a similar fashion with eq. (7) replaced by
\begin{equation}
g(\nu_1, \nu_2, T) =
\int_{\nu_1}^{\nu_2} d\nu~ \frac{L_{\nu}^{~M}(\nu, T)}{L_X^{~~M}}~
e^{-\beta}~.
\label{9}
\end{equation}

What about $\sigma_X(E)$? 
It is well known that for X-rays incoming on gas 
with interstellar composition 
$\sigma_X(E) \simeq 2.6 \times 10^{-22} E^{-8/3}$ 
cm$^{-2}$ with $E$ in keV (Morrison and McCammon \cite{mm}). 
On the other hand, in the limiting case of $Z=0$ it turns out
that $\sigma_X(E)\simeq 4 \times 10^{-24} E^{-3}$ cm$^{-2}$ 
(see Fig. 1 in Morrison and McCammon \cite{mm}).
Since we are dealing with the case $Z\sim 10^{-2}Z_{\odot}$, we shall use
for $\sigma_X(E)$ in eq. (\ref{4}) the linear interpolation between the
above expressions.

We report in Table 1 the values of $g(\nu_1,\nu_2,T)$ for various
energy bands and different values of the MACHO coronal gas 
temperature $T$ in both cases ($a$) and ($b$). As we can see, the fraction
of X-rays surviving the absorption by molecular clouds in DCs ranges
from $\sim 0.1\%$ to $\sim 74\%$. 
   \begin{table}
      \caption{
The attenuation function $g$ - see eqs. (\ref{8})-(\ref{9})  - 
is displayed for selected 
values of the MACHO coronal temperature $T$ in different energy bands
(the first and second ones are the ROSAT medium and high-energy bands
while the third refers to the EPIC instrument on the planned satellite
XMM). 
We consider two cases: ($a$) MACHOs uniformly distributed in the DCs 
and ($b$) MACHOs concentrated in the DC cores.
For the latter case, we give in the last column the X-ray 
flux $\Phi_X^{~~DC}$ - see eq.(\ref{12}) -
reaching the Earth, 
assuming, as an illustration, $f=1/2$, $r = 20$ kpc, $M_{DC} = 10^5 M_{\odot}$
and $L_X^{~~M} = 10^{27}$ erg s$^{-1}$.
We have verified that in the low ROSAT energy band 0.15-0.28 keV 
the X-ray absorption is so efficient that we cannot hope to detect any 
radiation coming from a DC. }
\label{Table1}
\begin{center}
\begin{tabular}{|c|c|c|c|c|}
\hline
$E_1-E_2   $ &$   T$& $g(\nu_1,\nu_2,T)$& $g(\nu_1,\nu_2,T)$& 
$\Phi_X^{~~DC}     $ \\
 (keV)       &    (K)  &                   &                   
& (erg cm$^{-2}$ s$^{-1}$)     \\
\hline
             &      &  case ($a$)       &  case ($b$)      &   case ($b$)        \\
\hline
$0.5  - 0.9$ &$10^6$&$4.4\times 10^{-3}$&$5.6\times 10^{-4}$& $5.8\times 10^{-18}$\\
             &$10^7$&$5.7\times 10^{-2}$&$1.8\times 10^{-2}$& $1.8\times 10^{-16}$\\
             &$10^8$&$2.1\times 10^{-2}$&$6.6\times 10^{-3}$& $6.8\times 10^{-17}$\\
\hline
$0.1  - 2.5$ &$10^6$&$1.7\times 10^{-2}$&$8.5\times 10^{-4}$& $8.8\times 10^{-18}$\\
             &$10^7$&$3.0\times 10^{-1}$&$2.2\times 10^{-1}$& $2.3\times 10^{-15}$\\
             &$10^8$&$2.3\times 10^{-1}$&$1.9\times 10^{-1}$& $2.0\times 10^{-15}$\\
\hline
$ 0.1 -10  $ &$10^6$&$1.5\times 10^{-1}$&$8.5\times 10^{-3}$& $8.6\times 10^{-17}$\\
             &$10^7$&$3.7\times 10^{-1}$&$2.6\times 10^{-1}$& $2.7\times 10^{-15}$\\
             &$10^8$&$7.4\times 10^{-1}$&$6.9\times 10^{-1}$& $7.3\times 10^{-15}$\\
\hline
\end{tabular}
\end{center}
   \end{table}

At this point, we take into account the absorption by the galactic disk.
Actually, the foregoing analysis can be easily extended
to this case by replacing $\tau_{DC}(\nu,b,y)$ with
$\tau_{DC}(\nu,b,y)+\tau_{\rm disk}(\nu)$. Although the latter quantity
depends on the galactic coordinates, for simplicity we take
the average value  
$\tau_{\rm disk}(\nu)\simeq 6\times 10^{19}\sigma_X(\nu)$ cm$^{-2}$,
\footnote{We use this value, which is slightly less than 
the usual one, since we are mainly interested in observations 
towards regions of very low $HI$ column density.} and so  
the flux from a single DC at distance $r$ from Earth is
\begin{equation}
\Phi_X^{~~DC}(\nu_1, \nu_2) =0.8~\displaystyle{\frac{f}{r^2}} 
\left(\frac{M_{DC}}{M_{\odot}}\right) 
L_X^{~~M}~g(\nu_1,\nu_2,T)~.
\label{12}
\end{equation}
Values of $\Phi_X^{~~DC}(\nu_1,\nu_2)$
are reported in Table 1, for the typical situation $r=20$ kpc,
$M_{DC}=10^5~M_{\odot}$, $f=0.5$ and $L_X^{~~M}=10^{27}$ erg s$^{-1}$.
Of course, this is only an illustrative case, since we expect 
$3\times 10^2~M_{\odot}\ut < M_{DC}\ut < 10^6~M_{\odot}$ and 
$r\ut >10-20$ kpc.

Let us now estimate the number ${\cal N}$ of DCs  within a square
degree in the sky. 
Taking indicatively $\sim 10^{12}~M_{\odot}$ for the 
mass of the dark halo 
of our galaxy (Zaritsky \cite{zaritsky}), 
we get ${\cal N}\simeq 2.4\times 10^7~(M_{\odot}/M_{DC})$ deg$^{-2}$.
Thus, for
$3\times 10^2~M_{\odot}\ut < M_{DC}\ut < 10^6~M_{\odot}$,
we expect ${\cal N}$ in the range $24-80,000$ deg$^{-2}$. 
On the other hand, the angular size of a DC at distance $r$ from Earth is 
$\theta\simeq 0.4\arcmin (M_{DC}/M_{\odot})^{1/3}
({\rm kpc}/r)$,
and so we get
$2.7\arcmin~({\rm kpc}/r)\ut < \theta\ut < 40\arcmin~({\rm kpc}/r)$
for $M_{DC}$ in the above range.
However, it is crucial to stress that the effective angular size of DCs
in the
X-ray band can be much smaller than that. Indeed, the DCs with
$3\times 10^2~M_{\odot}\ut < M_{DC}\ut < 5\times 10^4~M_{\odot}$ should have 
started core collapse, and so MACHOs are expected to be concentrated towards 
the centre.

An important issue concerns the spectrum of X-rays surviving absorption by 
molecular clouds and thereby emerging from a DC. 
The energy-dependence of $\sigma_X(E)$ implies (due 
to the more efficient X-ray absorption at low-energy) a hardening of the 
X-ray spectrum. In order to quantify this effect,
it is useful to define the hardness ratio $HR=(H-S)/(H+S)$ 
as in Hasinger et al. \cite{hasinger}, where $H$ and $S$ are the fluxes in the 
$0.1-0.4$ and $0.4-2.4$ keV bands, respectively. 
It turns out that absorption substantially increases 
the hardness ratio, leading to $HR \sim 1$.

\section{Observational implications}
   \begin{figure}[htbp]
   \vspace{8.0cm}
\includegraphics{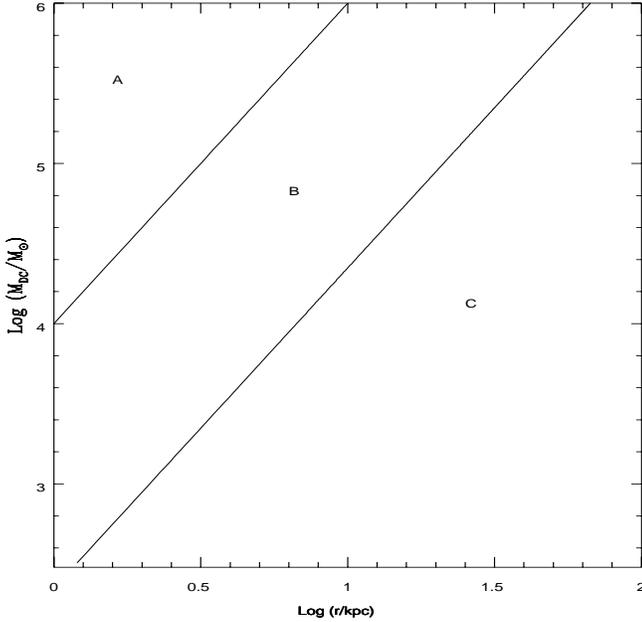} 
      \caption{
It is shown how the plane $M_{DC}-r$ gets divided into 
three regions depending on different observational expectations. 
Assuming a disk gas column density of $6\times 10^{19}$ cm$^{-2}$ (the same as
towards the {\it Lockman Hole}), DCs in region $A$ are observable as 
resolved sources.  
DCs in region $B$ are to show up as background fluctuations in 
deep-field exposures.  DCs in region $C$ can only 
contribute to the diffuse XRB.
              }
         \label{Figure1}
   \end{figure}
We turn now our attention to the observational implications of the above 
analysis. Various parameters -- namely $f$, $M_{DC}$, $T$ and $L_X^{~~M}$ -- 
play a key-r\^ole in the present considerations. 
Unfortunately, they are poorly constrained from a theoretical point of view.
Therefore, only by explicitly resorting to the experimental setup 
can we figure out under what conditions DCs are observable in the X-ray band. 
In order to make our discussion more definite, we shall
henceforth consider the typical situation in which $f\simeq 0.5$,
$T\simeq 10^7$ K and $L_X^{~~M}\simeq 10^{27}$ erg s$^{-1}$, whereas $M_{DC}$
will be left as a free parameter in the above-considered range (however,  
the results corresponding to $T\simeq 10^6$ K and 
$T\simeq 10^8$ K are exhibited in Table 1 and Fig. 2).
It should be pointed out that it still remains to be decided between cases
($a$) and ($b$) discussed in Section 2. As it can be seen from Table 1,
the difference turns out to be very small, and since realistically 
we expect an intermediate situation to occur,
we take for $g(\nu_1,\nu_2,T)$ the average value.

Presently, ROSAT is by far the most sensitive
X-ray telescope in the $0.1-2.5$ keV energy range, and so the subsequent
analysis will rest upon its instrumental capabilities. 
Its sensitivity threshold for detecting DCs as resolved sources
in medium-exposures is
$\simeq 10^{-13}$ erg cm$^{-2}$ s$^{-1}$, whereas a statistical analysis of the
{\it background fluctuations} in deep-exposures can be carried out down to 
$\simeq 2.5\times 10^{-15}$
erg cm$^{-2}$ s$^{-1}$ (Hasinger et al. \cite{hasinger}).
Combining these sensitivity thresholds with eq. (\ref{12}),  we single out the 
various observational prospects for the DCs.
More specifically (as it is shown in Fig. 1), the $r-M_{DC}$ parameter
plane gets divided into three regions $A,~ B$ and $C$ 
(notice that we also take into account the constraint  
$3\times 10^2~M_{\odot}\ut < M_{DC}\ut < 10^6~M_{\odot}$).
Region $A$ comprises DCs which are observable as resolved sources --
this region turns out to be severely constrained by ROSAT data. DCs in 
region $B$ are to show up as  background fluctuations 
in deep-field exposures. Finally, DCs in region $C$ are too faint
to be detected nowadays and would, therefore, merely contribute to the 
diffuse XRB.

\subsection{Dark clusters as resolved  objects } 
Let us first address the possibility of detecting DCs as resolved 
(discrete) sources. Inspection of eq. (\ref{12}) and Table 1 shows that 
the chance is presently rather dim. However, a 
considerable improvement is offered by the next-generation of 
satellite-borne X-ray detectors (like XMM), whose sensitivity is 
expected to be at least one order of magnitude better than for ROSAT.
Then the more massive DCs (with $M_{DC} \sim 10^6~M_{\odot}$) should be 
observed as resolved sources.
Moreover, it seems to us that the most promising strategy should be to look
towards a previously microlensed star in the LMC.

\subsection{Dark clusters and the new population of faint  X-ray sources}
Next, we proceed to discuss whether DCs form a new population 
of faint X-ray sources. 
Actually, fluctuation analyses of the XRB
in the direction of the {\it Lockman Hole} (with the lowest $HI$
column density) have led 
to the discovery of $\simeq 120$ deg$^{-2}$ unidentified discrete
sources in the $0.5-2$ keV energy band
with flux in the range 
$2.5\times 10^{-15}-10^{-13}$ erg cm$^{-2}$ s$^{-1}$
over the extrapolation from AGN evolution models 
(Hasinger et al. \cite{hasinger}).
As pointed out by these authors, one way to solve this discrepancy could 
be to adopt different parameters or more complicated evolutionary models 
for AGNs. Another explanation would be the existence of a new population 
of faint X-ray sources, which cannot be identified with any class of known 
objects (as stars, BL Lac objects, clusters of galaxies and normal galaxies).
A re-analysis of the same data (Hasinger \cite{hasinger_compton})
shows that most of these sources in excess (with respect to AGNs) 
may be identified with clusters of galaxies and narrow 
emission or absorption line galaxies.
Similar results have been obtained by analyses 
(down to a flux limit of 
$1.6 \times 10^{-15}$ erg cm$^{-2}$ s$^{-1}$) of ROSAT 
observations towards a different region in the sky with the deepest optically 
identified X-ray survey made so far (McHardy et al. \cite{mch}).
These observations further support 
the existence of an unidentified population of faint
(with flux $<3 \times 10^{-15}$ erg cm$^{-2}$ s$^{-1}$), point-like
(namely with angular size less than the instrumental resolution 
$\sim 20\arcsec$) X-ray sources. 

What is not clear at the moment is whether these sources are galactic or 
extragalactic in origin. Maoz and Grindlay \cite{maoz} have argued that the 
new population is wholly galactic. 

Our main point is that DCs can contribute to this new population of 
faint X-ray sources (of course, we are supposing that MACHOs are concentrated 
inside the DC cores). 
To this end, we have to compute the number ${\cal N}$ 
of DCs per square degree with flux above 
$2.5\times 10^{-15}$ erg cm$^{-2}$ s$^{-1}$ in the 
direction of the {\it Lockman Hole} (RA=$10^h~52^m~00^s$, $\delta=57^0~21
\arcmin~36\arcsec$).
Following Maoz and Grindlay \cite{maoz}, we first evaluate the distance
$r_*$ from Earth at which a DC with luminosity ${\cal L}_X^{~DC}$
yields a flux of
$2.5\times 10^{-15}$ erg cm$^{-2}$ s$^{-1}$. We find
$r_*\simeq 7\times 10^{-2}~(M_{DC}/M_{\odot})^{1/2}
~{\rm kpc}$.
Next, we assume for simplicity
that DCs are distributed according to the standard spherical
halo model. Denoting by $R$ the galactocentric distance
of a DC, the DC number density $n_{DC}(R)$ is 
\begin{equation}
n_{DC}(R) = n_0 \left(\frac{a^2+R_0^2}{a^2+R^2}\right)~,
\label{dmd}
\end{equation}
with 
$n_0 \simeq 0.9 \times 10^{-2}~(M_{\odot}/M_{DC})$ pc$^{-3}$, the core radius
$a \simeq 5$ kpc and our galactocentric distance $R_0 \simeq 8.5$ kpc. 
For our purposes, we parametrize a generic point 
in the halo by galactic spherical coordinates 
($r$, $b$, $l$) taking the Earth as the origin.
Accordingly, the relation between $r$ and $R$ is
$R(r) = (r^2+R_0^2-2rR_0 \cos b \cos l)^{1/2}$.
With eq. (\ref{dmd}), by straightforward steps we obtain 
\begin{equation}
{\cal N}\simeq 3\times 10^{-4}~n_0~ \int_{r_1}^{r_*}dr~ r^2
\frac{a^2+R_0^2}{a^2+R(r)^2}~~~~~~~~~~{\rm deg}^{-2}, \label{N}
\end{equation}
where the integral is taken along the line of sight towards the
{\it Lockman Hole} and $r_1\simeq 2.5$ kpc (corresponding to $R\simeq 10$ kpc).
In Table 2 we report ${\cal N}$ for
different values of $M_{DC}$, along with the total number ${\cal N}_T$
of DCs per square degree.
As it can be seen ${\cal N} \sim 20$ deg$^{-2}$ (for $M_{DC} \sim 
10^5~M_{\odot}$) and thus the DCs can explain at most 
$\sim$ 20\% of the new population of faint X-ray sources.
   \begin{table}
      \caption{
The number ${\cal N}$ of faint X-ray sources per square degree
towards the {\it Lockman Hole} with flux larger than 
$2.5 \times 10^{-15}$ erg cm$^{-2}$ s$^{-1}$ is shown  
in column 3 for some values of $M_{DC}$, 
assuming $L_X^{~~M} \sim 10^{27}$ erg s$^{-1}$.
The last column gives the total number of DCs ${\cal N}_T$ per square degree, 
irrespective of their flux (here we assume an amount of 
galactic dark matter of $\sim 10^{12}~M_{\odot}$).
}
         \label{Table2}
\begin{center}
\begin{tabular}{|c|c|c|c|}
\hline
$M_{DC}~(M_{\odot})$ & $n_0~({\rm pc}^{-3})$  & ${\cal N}$ & ${\cal N}_T$\\
\hline
$10^3$ & $9.3\times 10^{-6}$ & -- & $2.6\times 10^4$\\
$10^4$ & $9.3\times 10^{-7}$ & 18 & $2.6\times 10^3$\\
$10^5$ & $9.3\times 10^{-8}$ & 21 & $2.5\times 10^2$\\
$10^6$ & $9.3\times 10^{-9}$ & 12 & $2.6\times 10^1$\\
\hline
\end{tabular}
\end{center}
   \end{table}

Finally, two points have to be mentioned. 
First,   
the $\log N-\log S$ diagram (see Fig. 8 in Hasinger et al. 
\cite{hasinger}) shows a moderate deficit
of sources below a few  $10^{-15}$ erg cm$^{-2}$
s$^{-1}$ (which is roughly the flux expected from DCs of mass 
$\sim 10^5~M_{\odot}$ at a distance larger than $\sim 10-20$ kpc) 
with respect to the ${-3/2}$ behaviour 
(homogeneous distribution of X-ray sources). 
This fact can be 
naturally accounted for within the present model, 
since the source number density $n_{DC}(R)$ 
decreases as $R^{-2}$ at large distance, implying a flattening in the 
$\log N-\log S$ diagram from ${-3/2}$ to ${-1/2}$. 
However, it is difficult to disentangle the individual contributions 
from objects with a different distribution in space (see Fig. 4 in Hasinger 
\cite{hasinger_compton} and Fig. 6 in McHardy et al. \cite{mch}). 
Second, 
the average spectrum of the faint sources gets harder ($HR \rightarrow 1$)
at the lowest limiting flux (see Fig. 3 in Hasinger et al. \cite{hasinger}).
Also this fact can be explained within our model, since  
absorption (from molecular gas in DCs) reduces the X-ray flux on Earth
and increases $HR$.

\subsection{Dark clusters and the diffuse XRB}
Let us estimate the contribution of DCs to the diffuse XRB -- namely
the case for the DCs corresponding to region $C$ of Fig. 1. 
The expected X-ray flux per unit solid angle in the 
direction ($b,l$) is given by (in erg~cm$^{-2}$~s$^{-1}$~sr$^{-1}$)
\begin{equation}
\Phi_X( \nu_1, \nu_2; b,l) = 
\int_{r_{\rm min}(b,l)}^{r_{\rm max}(b,l)}dr ~n_{DC}(R(r))~
\frac{{\cal L}_X^{~~DC}(\nu_1,\nu_2)}
{4\pi} ~.
\label{fondo}
\end{equation}  
The best chance to detect the X-rays in question is provided by 
observations at high galactic latitude
(or towards the {\it Lockman Hole}) 
and thus we evaluate $\Phi_X(\nu_1,\nu_2;90^{0})$.
Therefore, we can safely take 
$r_{\rm min} \simeq 10$ kpc and $r_{\rm max}\simeq 100$ kpc.
The resulting X-ray fluxes are plotted in Figs. 2a - 2c
(for different energy bands) as a function of  
$T$ and for various galactic models.
For each band, the horizontal line shows, for comparison, the diffuse 
unresolved XRB excess
(with respect to the resolved discrete source contribution).
   \begin{figure}[htbp]
   \vspace{11.6cm}
  \includegraphics{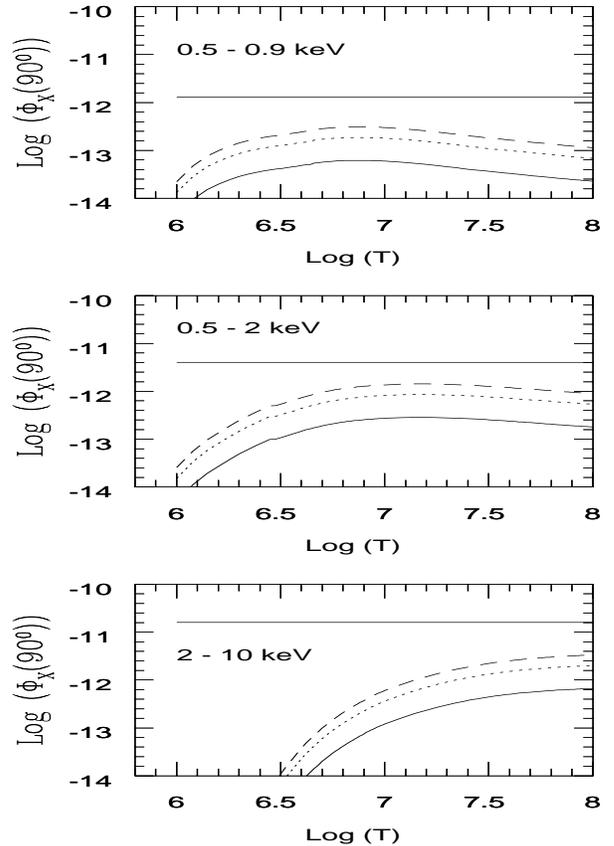}
      \caption{
The X-ray flux $\Phi_X(90^0)$ 
(in units of erg s$^{-1}$ cm$^{-2}$ deg$^{-2}$)
at high galactic latitude 
due to the DC emission - see eq. (\ref{fondo}) - is given, in various bands,
as a function of the MACHO temperature $T$ in K. To be definite we choose
$L_X^{~~M}\sim 10^{27}$ erg $s^{-1}$, $f\sim 0.5$ and an intermediate value
for $g$ (see Table 1).
We consider three different values for the total galactic dark matter 
$M_{\rm gal}$: 
the full lines are for $M_{\rm gal}= 1 \times 10^{12} ~ M_{\odot}$,
the dotted lines   for $M_{\rm gal}=3 \times 10^{12} ~ M_{\odot}$ and
the dashed lines   for $M_{\rm gal}=5 \times 10^{12} ~ M_{\odot}$.
For comparison, we show the corresponding XRB values in the various energy 
bands: $1.3\times 10^{-12}$, $4\times 10^{-12}$, $1.6\times 10^{-11}$ 
erg cm$^{-2}$ s$^{-1}$ deg$^{-2}$, respectively.
}
         \label{Figure2}
    \end{figure}
As we can see, the DC emission can at most account for a few percents of the
observed XRB flux.
A somewhat higher flux has been obtained by Kashyap et al. \cite{kashyap}
for an unclustered population of MACHOs.

\section{Limits on the Amount of Virialized Diffuse Gas}
One of the most intriguing recent results in the study of the 
diffuse XRB has been the detection by ROSAT of shadows in the 1/4 keV XRB
$0.15-0.28$ keV towards high-latitude interstellar clouds in 
Draco
(Burrows and Mendenhall \cite{burrows}, Snowden et al. \cite{snowden91}).
Data show that about one half of the emission originates beyond these
clouds and this fact is consistent with the presence of a 
$T\sim 10^6$ K diffuse halo gas in our galaxy
\footnote{
Other spirals are observed to have diffuse X-ray 
halos as a consequence of the presence of hot diffuse gas, heated by the 
galactic gravitational field at virial temperatures $\sim 10^6$ K 
(Cui et al. \cite{cui}). 
The inferred typical X-ray luminosities are 
$\sim 10^{39}-10^{40}$ erg s$^{-1}$,
substantially less than that of bright ellipticals.}.
Below, we 
are going to derive an upper bound on the amount of this virialized 
diffuse halo gas 
by fully tracing back to it the whole XRB excess in the $0.5-2$ keV band.
Assuming that a fraction $\eta$ of the galactic halo dark matter is in the 
form of virialized diffuse gas and that this is distributed 
like the dark matter 
(namely according to eq. (\ref{dmd})), the X-ray flux on Earth 
(in units of erg~cm$^{-2}$~s$^{-1}$~sr$^{-1}$) is
\begin{equation}
\Phi_X^{~~{\rm gas}}(b,l)=\frac{\eta^2}{4\pi}
\left(\frac{\rho_0}{\mu m_H}\right)^2
I_1(b,l)~I_X(T,Z,b,l) ~,
\label{dhg}
\end{equation}
where $\mu \sim 1.22$ is mean molecular weight of the ionized gas, 
$\rho_0 \simeq 0.015~M_{\odot}$ pc$^{-3}$ is the local dark matter density, 
while $I_1(b,l)$ and $I_X(T,Z,b,l)$ are as follows
\begin{equation}
I_1(b,l)=\int_0^{r_{\rm max}} dr
\left(\frac{a^2+R_0^2}{a^2+R(r)^2}\right)^2~,
\end{equation}
\begin{equation}
\begin{array}{ll}
I_X(T,Z,b,l) =
\int_{\nu_1}^{\nu_2} d\nu~f(\nu,T,Z)~e^{-\tau_{\rm disk}(\nu,b,l)}~,
\end{array}
\end{equation}
where $f(\nu,T,Z)$ is defined via $\epsilon_{\nu}$ and
$\tau_{\rm disk}(\nu,b,l)$ denotes the optical depth of the disk.
Table 3 gives $\eta$ as a function of $T$ for several values of 
$Z$. As we can see, no more than $5\%$ of the matter in the halo of
our galaxy can be in the form of virialized diffuse gas.
This conclusion makes the possibility of an {\it unclustered} MACHO
population in the galactic halo unplausible, since then the leftover
gas would remain diffuse within the halo and should be virialized today.
   \begin{table}
      \caption{
The upper limit on the fraction $\eta$ of halo dark matter in the form 
of diffuse gas is given for some values of the gas temperature $T$ 
(close to the virial temperature $1.6 \times 10^6$ K)
and for some values (arbitrarily selected) of gas metallicity $Z$.
The fraction $\eta$ is evaluated by means of eq. (\ref{dhg})
assuming that the whole XRB excess at high galactic latitude is due to 
the diffuse gas emission. 
We consider a total mass and radius 
of the Galaxy $\sim 1.1 \times 10^{12}~M_{\odot}$ and
$\sim 120$ kpc, respectively. 
}
         \label{Table3}
\begin{center}
\begin{tabular}{|c|c|c|c|}
\hline
$T$ (K) $=$      &$1.2 \times 10^6$ & $1.6 \times 10^6$& $2.0 \times 10^6$\\
\hline
$Z=10^{-3}~Z_{\odot}$      & 5.2             & 3.3              & 2.5   \\
$Z=10^{-2}~Z_{\odot}$      & 4.9             & 3.1              & 2.4   \\
$Z=10^{-1}~Z_{\odot}$      & 3.1             & 2.0              & 1.6   \\
$Z=1 ~~~~~~Z_{\odot}$      & 1.2             & 0.8              & 0.6  \\
\hline
\end{tabular}
\end{center}
   \end{table}

\section{Discussion and conclusions}
The possibility that MACHOs with mass $\sim 0.1~M_{\odot}$ 
possess, in analogy with low-mass stars, coronal X-ray emission has several 
observational consequences. Indeed, according to the considered 
model for the halo dark
matter, the large number of MACHOs inside each DC
gives rise to a total X-ray luminosity ${\cal L}_X^{~~DC}$ as high as 
$10^{29}-10^{32}$ erg s$^{-1}$, in spite of the strong 
absorption brought about by molecular clouds.

The main parameter discriminating between different 
values of ${\cal L}_X^{~~DC}$ is $M_{DC}$ (see Fig. 1), which lies in 
the broad range $3 \times 10^2- 10^6~M_{\odot}$.
At the lower end of this range, namely for 
$M_{DC} \ut < 10^4~M_{\odot}$, DCs can 
contribute to the XRB at a level of at most
5\%. Of course,  the latter value depends on 
the assumed total amount of halo dark matter, the coronal MACHO temperature 
and the considered energy range. 
For $M_{DC} \sim 10^5~M_{\odot}$
we get a different observational situation. In this case, fluctuation
analyses of the XRB should yield $\sim 20$ DCs per square degree
with X-ray flux in the range 
$2.5\times 10^{-15}-10^{-13}$ erg cm$^{-2}$ s$^{-1}$.
This conclusion supports the idea that
DCs contribute to the new population of faint
X-ray sources advocated by Hasinger et al. \cite{hasinger,hasinger_compton} 
and  McHardy et al. \cite{mch}. 
Finally, if $M_{DC}\sim 10^6~M_{\odot}$, some DCs should be
observable as resolved sources with the future planned satellite missions.
The best strategy would then be to look in the direction of a previously
microlensed star towards the LMC.

We have also derived an upper limit on the amount of virialized diffuse
halo gas by using ROSAT data.  Even if dependent on the 
gas virial temperature (and thus on the total dynamical mass of 
the Galaxy) and on the metallicity, this limit shows that the amount of 
diffuse gas in the galactic halo is less than 5\%.
This result corroborates our assumption that the primordial gas 
left over from MACHO formation should remain
within the DCs (in the form of cold molecular clouds), for otherwise
it would be heated by the gravitational field thereby emitting in the 
X-ray band.
 
\acknowledgements{
We would like to thank A. Fabian for useful discussions. 
FDP is partially supported by the {\em Dr. Tomalla Foundation}, GI by the 
{\em Agenzia Spaziale Italiana} and MR by the {\em Dipartimento di Fisica 
Nucleare e Teorica, Universit\`a di Pavia}.}


\begin{thebibliography}{}
\bibitem[1993]{alcock}
Alcock, C. et al. 1993, Nat 365, 621
\bibitem[1993]{aubourg}
Aubourg, E. et al. 1993, Nat 365, 623
\bibitem[1993]{barbera}
Barbera, M. et al. 1993 ApJ 414, 846
\bibitem[1991]{burrows}
Burrows, D. N. \& Mendenhall, J. A. 1991, Nat 351, 629
\bibitem[1996]{cui}
Cui, W. et al. 1996, ApJ 468, 102
\bibitem[1995]{depaolis2}
De Paolis, F., Ingrosso, G., Jetzer, Ph. \& Roncadelli, M. 1995, A\&A 
295, 567
\bibitem[1996]{dij}
De Paolis, F., Ingrosso, G. \& Jetzer, Ph. 1996, ApJ 470, 493
\bibitem[1996a]{vangelo}
De Paolis, F., Ingrosso, G., Jetzer, Ph. \& Roncadelli, M. 1996a, Int. J.
of Mod. Phys.D 5, 151 
\bibitem[1996b]{apocalisse}
De Paolis, F., Ingrosso, G., Jetzer, Ph. \& Roncadelli, M. 1996b, 
preprint ZU-TH 8/96, submitted to ApJ
\bibitem[1996]{drake}
Drake, J. J. et al. 1996, ApJ 469, 828
\bibitem[1996]{evans} 
Evans, N. W. 1996, astro-ph 9611161
\bibitem[1985]{fall}
Fall, S. M. \& Rees, M. J. 1985, ApJ 298, 18
\bibitem[1993]{fleming}
Fleming et al. 1993, ApJ 410, 387
\bibitem[1996]{ggt}
Gates E.J., Gyuk G. \& Turner M., 1996, Phys. Rev., D53, 4138 
\bibitem[1996]{gerhard}
Gerhard O. \& Silk J. 1996, ApJ 472, 34
\bibitem[1993]{hasinger}
Hasinger, G et al. 1993, A\&A 275, 1
\bibitem[1996]{hasinger_compton}
Hasinger, G. 1996, A\&AS 120, 607
\bibitem[1994]{kashyap}
Kashyap, V.,  Rosner, R., Schramm, D. \& Truran, J. 1994, ApJ 431, L87
\bibitem[1995]{maoz}
Maoz, E \& Grindlay, J. E. 1995, ApJ 444, 183
\bibitem[1997]{mch}
McHardy, I. M. et al. 1997, astro-ph 9703163
\bibitem[1983]{mm}
Morrison, R. \& McCammon, D. 1983, ApJ 270, 119
\bibitem[1996]{mullan}
Mullan, D. J. \& Fleming, T. A. 1996, ApJ 464, 890
\bibitem[1997]{nulfab}
Nulsen P.E.J. \& Fabian A.C., 1997, submitted
\bibitem[1977]{rs}
Raymond, J. C. \& Smith, B. W. 1977, ApJS 35, 419
\bibitem[1990]{schmitt}
Schmitt, J. H. M. M. et al. 1990, ApJ 365, 704
\bibitem[1991]{snowden91}
Snowden, S. L. et al. 1991, Science 252, 1529 
\bibitem[1989]{zaritsky}
Zaritsky, D. et al. 1989, ApJ 345, 759
\end{thebibliography}
\end{document}